\documentclass[%
aip,
 amsmath,amssymb,
 reprint,%
]{revtex4-1}

\usepackage{graphicx}
\usepackage{dcolumn}
\usepackage{bm}
\usepackage[utf8]{inputenc}
\usepackage[T1]{fontenc}
\usepackage{mathptmx}
\usepackage{etoolbox}

\makeatletter
\def\@email#1#2{%
 \endgroup
 \patchcmd{\titleblock@produce}
  {\frontmatter@RRAPformat}
  {\frontmatter@RRAPformat{\produce@RRAP{*#1\href{mailto:#2}{#2}}}\frontmatter@RRAPformat}
  {}{}
}%
\makeatother
\begin{document}

\preprint{AIP/123-QED}

\title{\textit{Time}SOAP:
Tracking high-dimensional fluctuations in complex molecular systems \textit{via} time-variations of SOAP spectra}

\author{Cristina Caruso}
 \affiliation{Department of Applied Science and Technology, Politecnico di
Torino, Corso Duca degli Abruzzi 24, 10129 Torino, Italy}

\author{Annalisa Cardellini}
\affiliation{Department of Innovative Technologies, University of Applied
Sciences and Arts of Southern Switzerland, Polo Universitario Lugano, Campus Est, Via la Santa 1, 6962 Lugano-Viganello, Switzerland}

\author{Martina Crippa}
\affiliation{Department of Applied Science and Technology, Politecnico di
Torino, Corso Duca degli Abruzzi 24, 10129 Torino, Italy}

\author{Daniele Rapetti}
\affiliation{Department of Applied Science and Technology, Politecnico di
Torino, Corso Duca degli Abruzzi 24, 10129 Torino, Italy}

\author{Giovanni M. Pavan}
\email{corresponding author: Giovanni M. Pavan
(giovanni.pavan@polito.it)}
\affiliation{Department of Applied Science and Technology, Politecnico di
Torino, Corso Duca degli Abruzzi 24, 10129 Torino, Italy}
\affiliation{Department of Innovative Technologies, University of Applied Sciences and Arts of Southern Switzerland, Polo Universitario Lugano, Campus Est, Via la Santa 1, 6962 Lugano-Viganello, Switzerland}

\date{\today}

\begin{abstract}
Many molecular systems and physical phenomena are controlled by local fluctuations and microscopic dynamical rearrangements of the constitutive interacting units that are often difficult to detect. This is the case, for example, of phase transitions, phase equilibria, nucleation events, and defect propagation, to mention a few. A detailed comprehension of local atomic environments and of their dynamic rearrangements is essential to understand such phenomena, and also to draw structure-property relationships useful to unveil how to control complex molecular systems. Considerable progresses in the development of advanced high-dimensional structural descriptors (\textit{e.g.}, Smooth Overlap of Atomic Position (SOAP), etc.) have certainly enhanced the representation of atomic-scale simulations data. However, despite such efforts, local dynamic environment rearrangements remain still difficult to elucidate. Here, exploiting the structural-rich description of atomic environments of SOAP and building on the concept of time-dependent local variations, we developed a time-dependent SOAP-based descriptor, \textit{Time}SOAP (\textit{$\tau$}SOAP), which essentially tracks the time variations in the local SOAP environments surrounding each molecule (\textit{i.e.}, each SOAP center) in complex molecular systems along ensemble trajectories. We demonstrate how analysis of the time-series \textit{$\tau$}SOAP data and of their time-derivatives allows detecting dynamics domains and tracking instantaneous changes of local atomic arrangements (\textit{i.e.}, local fluctuations) in a variety of molecular systems. The approach is simple and general, and we expect will help to shed light on a variety of complex dynamical phenomena.
\end{abstract}

\maketitle
\section{\label{sec:level1}INTRODUCTION}
Structure-property relationships, at the heart of modern materials science, are hard to be elucidated in complex molecular systems. Multi-scale and many-body interactions among all the atoms make it challenging, yet inspiring, to reconstruct the macroscopic behavior of such systems from the underlying atomic structure.\cite{chapman2022efficient, stavroglou2020unveiling} 
Ranging from materials with an intrinsically dynamic character such as soft supramolecular architectures, to common crystal lattices, a thorough knowledge of atomic arrangements, including their structural and dynamic evolution, is required to increasingly unlock tangled material responses and features.\cite{bochicchio2017into, bochicchio2019defects, albertazzi2014probing, wang2022structural, sosso2018unravelling} In crystalline solids, for instance, material's plasticity, viscosity, and microstructure's evolution are dictated by the energy and kinetics of defects\cite{goryaeva2020reinforcing}, or structural imperfections. \cite{lifshitz1966dynamics, dove1993introduction} Furthermore, atomically disordered domains such as surfaces, grain boundaries, and heterogeneous interfaces have been widely recognized to be linked to transport, mechanical, electronic, and optical
properties.\cite{stachurski2011structure, li2018recent, zhou2020thermal, yan2019research} Shedding light on the vital connection between the complex atomic arrangements and the material dynamic properties would clearly pave the way for novel design rules and optimization of molecular systems for tailored behaviors.\cite{rosenbrock2017discovering, cardellini2021modeling} However, although desirable, such material design objective meets a number of practical and theoretical challenges thus standing for the most ambitious goal in material science.

In recent years, the advances in data availability and computational power have enabled the development of valuable tools for gaining a deeper understanding into chemical-physical phenomena occurring in materials\cite{tantakitti2016energy, carter2019accurate} In particular, molecular dynamics (MD) simulations have been playing an increasingly significant role in the exploration of materials, providing a large source of potential information.\cite{frederix2018molecular, lee2012modeling, bejagam2015supramolecular, perego2021multiscale, bochicchio2017cooperative, behler2007generalized, bartok2010gaussian, crippa2022molecular, souza2021martini} 
The use of MD simulations to elucidate structure-property relationships substantially embeds two-steps level protocol: (i) the translation of MD trajectories into a numerical representation of atomic neighborhood environments, resulting in high-detailed and high-dimensional data, known as \textit{fingerprints} or \textit{descriptors}; (ii) the extrapolation of meaningful information from the large volumes of generated data sets. Regarding the latter step, Machine Learning (ML) algorithms have often revealed promising advantages to handle the large and complex set of data, thereby achieving increased interest. \cite{abrams2013enhanced, ferguson2017machine, butler2018machine, jackson2019recent, hase2020designing, glielmo2021unsupervised} However, a low-dimensional representation facilitating the navigation and identification of hidden patterns and features would be desirable. 

Within this framework, methods for adequately characterizing complex atomic arrangements from MD simulations have received a remarkable expansion. 
Over the last decades, many \textit{descriptors} have been proposed relying on the use of order parameters or mathematical quantities.\cite{chapman2022efficient} 
Low-dimensional descriptors based on the use of order parameters often allow to gain very accurate information, though being dependent from \textit{a priori} knowledge about systems' features. 
However, methods operating on structural environments (i.e., order parameters) such as the coordination analysis, bond order analysis, \cite{steinhardt1981point} bond angle analysis (BAA),\cite{ackland2006applications} common neighbor analysis, (CNA)\cite{honeycutt1987molecular} adaptive CNA (A-CNA),\cite{stukowski2012structure} and Voronoi analysis, generally struggle to identify different local coordination environments when the geometric symmetry is lost or exhibits a short-range nature (e.g. in crystalline systems close to the melting temperature). 
On the other hand, coupling more mathematically-sophisticated descriptors to ML approaches enables effective characterization of systems by exploiting the rich and high-dimensional data sets provided by MD simulations,\cite{cheng2020mapping, ceriotti2019unsupervised, wang2019machine} also being less dependent from \textit{a priori} knowledge. 
Nevertheless, advanced mathematical descriptors such as the Behler-Parrinello symmetry functions (BP),\cite{behler2011atom} Chebyshev polynomial representations (CPR),\cite{artrith2017efficient} adaptive generalizable neighborhood informed features (AGNI),\cite{batra2019general, chandrasekaran2019solving} smooth overlap of atomic positions, (SOAP)\cite{bartok2013representing} and atomic cluster expansion (ACE)\cite{drautz2019atomic} generally operate on atomic environments, that still represent local properties and weakly capture global and dynamics pictures.

Among more mathematically related descriptors, SOAP turned out to be very efficient in the characterization of a wide plethora of systems \cite{monserrat2020liquid,offei2022high,de2016comparing,reinhardt2020predicting} including soft disordered and complex assemblies.\cite{gasparotto2019identifying,cardellini2022,capelli2021data,lionello2022supramolecular} 
Despite being strongly connected to the structural features of local environments, the SOAP fingerprint coupled with unsupervised clustering approaches and statistical analyses has been recently used also to reconstruct the dynamics of complex systems such as, \textit{e.g.}, metal surfaces,\cite{cioni2022} metal nanoparticles,\cite{rapetti2022} soft supramolecular polymers,\cite{gasparotto2019identifying,gardin2022classifying} self-assembled micelles\cite{cardellini2022} and complex hierarchical superlattices, to cite a few.\cite{bian2021electrostatic,lionello2022supramolecular}
Since SOAP descriptors are typically high-dimensional, both linear and non-linear dimensionality reduction (DR) approaches are often employed for facilitating both analyses and data visualization.\cite{facco2017estimating, tenenbaum2000global, coifman2006diffusion, scholkopf1998nonlinear}
However, DR represents the fundamental roadblock because it inherently leads to a loss of information, resulting in a challenging characterization of systems where ordered and disordered domains coexist in dynamic exchange and equilibrium. 
In addition, beyond some valuable techniques,\cite{schwantes2015modeling, tsai2021sgoop} the time evolution of structural changes, including rare fluctuations, still remain weakly explored by simply classifying datasets with unsupervised and sophisticated ML tools.

Time-dependent descriptors offer a different approach. For example, a recently developed descriptor - Local Environments and Neighbors Shuffling (LENS)\cite{crippa2022} - monitors how much the microscopic surrounding of each molecular unit in a molecular system changes over time in terms of neighbor individuals along an MD trajectory, identifying dynamic domains and detecting local fluctuations in a variety of systems.\cite{crippa2022}
However, while such a descriptor keeps track of local dynamic rearrangements (\textit{e.g.}, local reshuffling, addition/subtraction of neighbors within a cutoff over time, etc.), it does not retain local structural information (\textit{e.g.}, relative displacement/adjustment of the neighbors within a cutoff over time), and it is thus not very well suited, \textit{e.g.}, to track local structural fluctuations and rearrangements.
A time-dependent descriptor capable of retaining rich structural information and monitoring efficiently structural changes over time would be desirable.

Building on such a concept, here we report a time-dependent descriptor, \textit{Time}SOAP (\textit{$\tau$}SOAP), which essentially exploits the structural-rich description of molecular/atomic environments guaranteed by the SOAP vectors and measures to what extent the SOAP power spectra of each unit in a complex molecular system changes over time. An ML-based analysis of the time-series \textit{$\tau$}SOAP data allows us to robustly and efficiently detect, \textit{e.g.}, structural transitions, phase transitions, and the coexistence of phases in a variety of systems with rich and diverse intrinsic dynamics.
Noteworthy, the time derivative of \textit{$\tau$}SOAP also provides sharp signals identifying local fluctuations, highlighting local and rare events that may be overlooked with other approaches.
The paper is organized as follows. In Section II (Methods), we present our \textit{$\tau$}SOAP and \textit{$\tau$}SOAP-based descriptors and the coupled ML-based workflow. In Section III, we discuss the results obtained by performing our \textit{$\tau$}SOAP analysis on various systems characterized by solid/liquid coexisting phases, solid-like and fluid-like behaviors, respectively.   
Our tests indicate that \textit{$\tau$}SOAP analyses are flexible and robust, and can shed light on complex molecular/atomic systems with non-trivial multilayered dynamics providing insights that are difficult to attain with other approaches.
 
\section{Methods}
\subsection{\label{sec:level2}SOAP as a descriptor of atomic environments}
Recently, data-driven approaches capturing the structural complexity of materials from equilibrium MD trajectories have been proposed. A generic MD trajectory is represented by an ordered list of N atomic coordinates \textbf{R}(t) in the 3D space at each simulation time step, where N is the number of particles in the system. In order to characterize complex atomic arrangements, descriptors of atomic neighborhood environments have been widely employed. By associating a \textit{feature} vector to each \textbf{R}(t), the descriptors enable to pass from the 3D \textit{coordinate} space to an S-dimensional \textit{feature} space. Importantly, these representations are required (1) to be permutationally, translationally and rotationally invariant, in order for physically equivalent configurations to be recognised as such, and (2) to smoothly vary with small changes in atomic positions. Among many developed descriptors, we adopt the Smooth Overlap of Atomic Position (SOAP) to examine our sample of materials ranging from crystalline to soft and liquid states. SOAP is a state-of-art, high-dimensional representation of atomic environments and it has recently provided valuable insights on both properties and structural features\cite{helfrecht2019atomic, gardin2022classifying, de2016comparing, willatt2019atom}.\\
The SOAP descriptor centers Gaussian density distributions on each atom. For a given atom, a smooth representation of the neighbor density is generated from the sum of Gaussians centered on each surrounding atom, namely:
\begin{equation}
\centering
   \rho^i(\textbf{r}) = \sum_{j} \exp\left[\frac{-|\textbf{r} - \textbf{r}_{ij}|^{2}}{2\sigma^{2}}\right] f_{rcut}(|\textbf{r}-\textbf{r}_{ij}|),
\label{eq:one}
\end{equation}

where to each neighbor center \textit{j}, located at a distance $\textbf{r}=\textbf{r}_{ij}$ from the \textit{i}-th center, a Gaussian function is associated. $\sigma$ is the distribution width of each Gaussian. The environment related to each center \textit{i} incorporates information up to a fixed cutoff, $\textit{rcut}$, where the function $\textit{f}_{rcut}$ smoothly goes to 0. Then, by expanding the Eq.~(\ref{eq:one}) in the basis of orthonormal radial functions $\textit{R}_\textit{n}(r)$ and spherical harmonics $\textit{Y}_\textit{l,m}(\textbf{\^{r}})$, the corresponding SOAP power-spectrum is calculated. For the \textit{i}-th center, it can be expressed as:
\begin{equation}
\centering
   \gamma^{i}_{nn'l} \propto \frac{1}{\sqrt{2l+1}} \sum^{+l}_{m=-l} (c^{i}_{nlm})*c^{i}_{n'lm},
\label{eq:two}
\end{equation}

with $c^{i}_{nlm}$ representing the expansion coefficients of the neighbor density associated to the \textit{i}-th center. The parameters \textit{n} and \textit{n'} range from 1 to \textit{nmax}, while \textit{l} index runs from 1 to \textit{lmax}. From the values of \textit{nmax} and \textit{lmax} it is possible to derive the dimension S of the full SOAP feature vector, which can be written as:
\begin{equation}
\centering
   \textbf{p}_{i} = \{\gamma^{i}_{nn'l}\},
\label{eq:three}
\end{equation}

representing the SOAP descriptor associated to the \textit{i}-th center, which includes all the contributions from the Eq.~(\ref{eq:two}).
Here, we used in-house code, SOAPify,\cite{SOAPify} to compute the SOAP vectors, with \textit{nmax}, \textit{lmax}=8, and different $\textit{rcut}$ values depending on the characteristics of the considered system. From the 3D coordinate vector corresponding to each MD simulation time, we calculate the SOAP vector $\textbf{p}_{i}$ for a selected set $\{i\}$ of centers (referred to as SOAP centers). In summary, we obtain a dataset containing S-dimensional SOAP vectors describing the structural arrangements related to the $\{i\}$ selected sites at each sampled configuration. Since these SOAP vectors encode the information about the atomic environments surrounding each center, SOAP is referred to as a "local" descriptor.

In order to evaluate how similar are two environments centered in two sites, a similarity measure has been defined by means of a linear kernel of their neighbor density representations: 
\begin{equation}
\centering
   K^{SOAP}(i,j) = (\textbf{q}_{i} \cdot \textbf{q}_{j}).
\label{eq:four}
\end{equation}

Since $\textbf{q}=\frac{\textbf{p}}{|\textbf{p}|}$, that is, the unit-normalized SOAP vector, $K^{SOAP}(i,j)$ goes from 0 for no overlapping to 1 for completely superimposed vectors. 
Furthermore, from Eq.~(\ref{eq:four}), a metric referred to as "SOAP distance" between two environments can be defined:
\begin{equation}
\centering
   d^{SOAP}(i,j) = \sqrt{2-2 \cdot K^{SOAP}(i,j)} \propto \sqrt{2-2\textbf{p}_{i}\textbf{p}_{j}} .
\label{eq:five}
\end{equation}

Importantly, $\textbf{p}_{i}$ and $\textbf{p}_{j}$ describe the local environments related to two \textit{different} SOAP \textit{centers}.
Besides the SOAP kernel, this distance representation provides a bounded similarity measure between two local environments, indicating how their local densities match in the S-dimensional feature space.

\begin{figure*}[ht!]
\centering
\includegraphics[width=18cm]{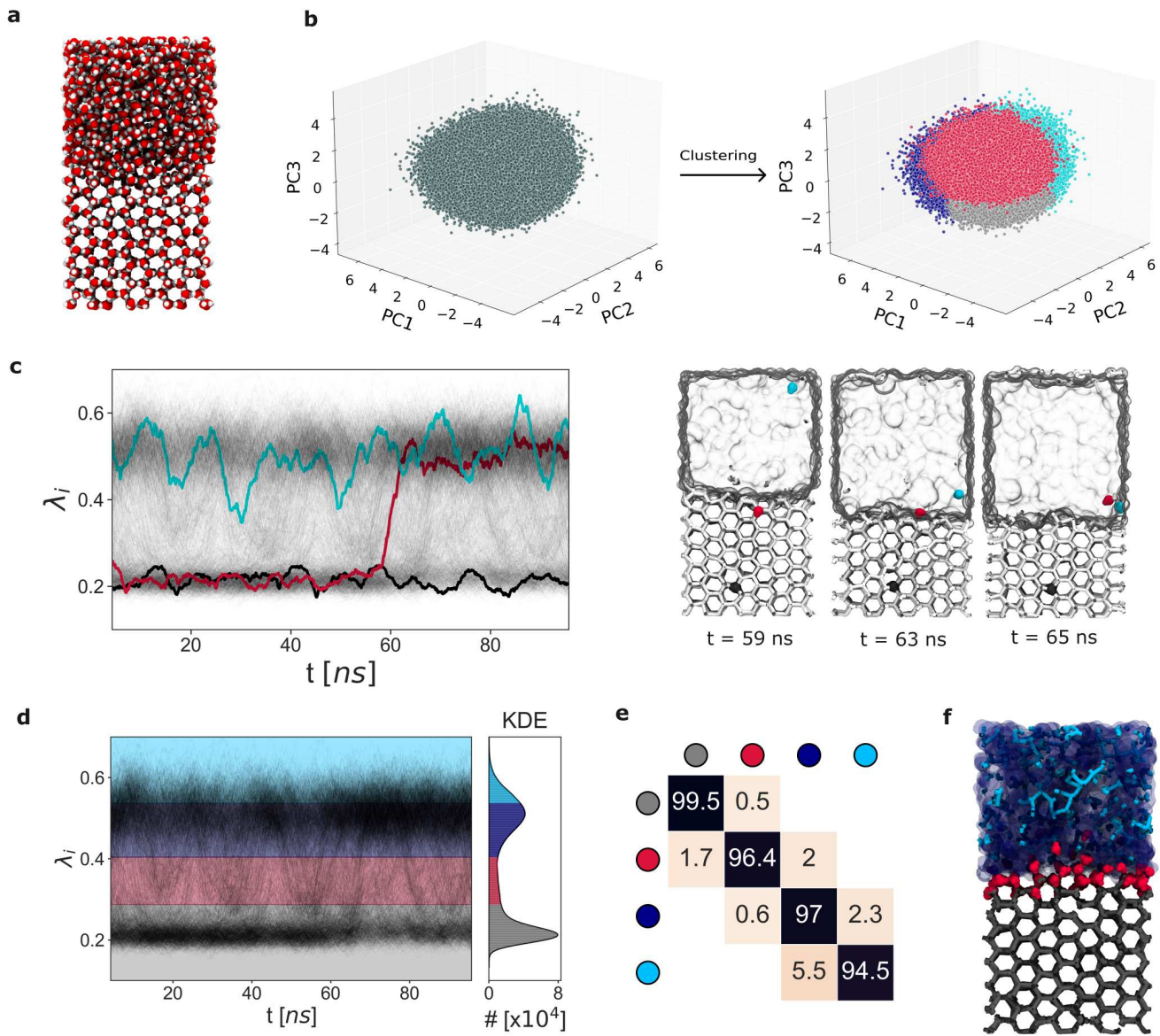}
\caption{Automatic detection of molecular motifs in ice/liquid water coexistence. (a) MD snapshot of ice/liquid water simulation box made of 2048 TIP4P/ICE water molecules at $T = 268$ K. Color code: red for oxygen and white for hydrogen atoms. (b) Example of a typical SOAP-based pattern recognition procedure. Left: PCA projection of the SOAP-based data-set estimated from the ice/liquid water system in (a). Right: clustering analysis - on the same data-set - carried out with KMeans. (c) On the left, time-series of \textit{$\tau$}SOAP signals, $\lambda_i(t)$, shown for all the oxygen atoms in (a). The colored $\lambda_i(t)$ profiles are related to three explicative oxygen atoms, i.e. (i) black, (ii) cyan, and (iii) crimson, displayed on the right with the respective color code. The reported MD snapshots are around $t \sim 60$ ns. (d) \textit{$\tau$} SOAP-based analysis. ${\lambda}_{i}(t)$ profiles and their KDE are carried out for all the oxygen atoms of all water molecules in the system (a). The final $k = 4$ detected macro-clusters are shown as colored in gray, crimson, blue, and cyan. (e) Interconnection probability matrix of the final $k = 4$ identified macro-clusters. (f) MD snapshot showing the four main clusters identified by \textit{$\tau$} SOAP-based analysis (same color code of (d)): ice (in gray), solid/liquid interface (in crimson), liquid water (in blue), a distinct domain in the liquid phase (in cyan).}
\label{fig:fig01}
\end{figure*}

\subsection{\label{sec:level2}Tracking dynamical SOAP variations with \textit{Time}SOAP}
The output dataset containing the S-dimensional SOAP vectors is typically high-dimensional, and although rich in information on the atomic/molecular arrangements, requires a crucial pre-processing to both facilitate the interpretation of the results and effectively identify relevant molecular patterns. For this reason, after estimating $\textbf{p}_{i}$ (Eq.~(\ref{eq:three})) for the whole set of SOAP centers $\{i\}$ at each sampled configuration of the MD trajectory, a SOAP-based pattern recognition procedure typically relies on two successive key phases: (1) use of a dimensionality reduction (DR) of SOAP spectra by means, for instance, Principal Component Analysis\cite{kpfrs1901edinburgh, hotelling1932edgeworth} (PCA); (2) employment of unsupervised clustering techniques for the identification of molecular motifs. To provide an example of this procedure, we consider a system where ice and liquid water coexist (Fig.~\ref{fig:fig01}a) at the melting/solidification temperature. Fig.~\ref{fig:fig01}b shows the SOAP fingerprints projected onto a 3-dimensional feature space \textit{via} PCA (1) and, then, the pattern recognition through unsupervised clustering (2). Despite providing some information on a wide range of molecular systems, this approach presents some key shortcomings: (i) since the time information is not emphasized, insights on consequential transition events as well as the temporal persistence of the individual molecular configurations are not retained, thus hindering a detailed comprehension of the rate of change of every individual molecular configuration; (ii) on such low-dimensional SOAP-based data set, some poorly-populated configurations may remain undetected by (\textit{e.g.}, density-based) unsupervised clustering approaches. This makes detecting local fluctuations and rare events typically awkward with such approaches. 

In this work, we propose an alternative procedure allowing to retain the time information from the high-dimensional SOAP vectors. Building upon the SOAP distance $d^{SOAP}(i,j)$ introduced above, we present a new SOAP-based fingerprint, named "\textit{Time}SOAP (\textit{$\tau$}SOAP)", which quantifies the local environment variation, over time, of each individual SOAP center \textit{i}. Indicating by $\lambda_{i}$ the variable form of \textit{$\tau$}SOAP, its instantaneous value is defined as:
\begin{equation}
\centering
   \lambda^{t+ \Delta t}_{i} = \frac{\sqrt{2-2 \cdot K^{SOAP}(i^{ t},i^{ t+ \Delta t})}}{\Delta t} \propto \frac{\sqrt{2-2\textbf{p}^{t}_{i}\textbf{p}^{t+ \Delta t}_{i}}}{\Delta t} .
\label{eq:six}
\end{equation}

Differently than the Eq.~(\ref{eq:five}), here both $\textbf{p}^{t}_{i}$ and $\textbf{p}^{t+ \Delta t}_{i}$ describe the local environments related to the \textit{same unit} (i.e., the \textit{i}-th SOAP center) but at \textit{different simulation times}, $t$ and $t+ \Delta t$, respectively. Thus, $\lambda^{t+ \Delta t}_{i}$ measures how similar the \textit{i}-th SOAP vector calculated at time $t$ is to that calculated at the next sampled timestep ($t+ \Delta t$). We analyze consecutive frames, namely adjacent points, where $\Delta t$ represents the MD sampling timestep (different for the various systems, see Molecular Dynamics Simulations for more details). As a result, \textit{$\tau$}SOAP evaluates how the \textit{i}-th local environment changes, in terms of SOAP descriptor, at every consecutive time interval $\Delta t$. We thus obtain ${\lambda}_{i}(t)$, namely a \textit{$\tau$}SOAP signal over time for each individual in the selected set $\{i\}$, thereby allowing to track the evolution of each SOAP constituent unit center along the trajectory. 

We can take a further step by estimating $\dot{\lambda}_{i}$, namely the first time-derivative of \textit{$\tau$}SOAP signal. Using the NumPy\cite{harris2020smith} Python package, we have:
\begin{equation}
\centering
   \dot{\lambda}^{t+ \Delta t}_{i} = \frac{\lambda^{t+ \Delta t}_{i} - \lambda^{t}_{i}}{\Delta t}.
\label{eq:seven}
\end{equation}

By computing it along the MD trajectory, we get $\dot{\lambda}_{i}(t)$. What $\dot{\lambda}_{i}$ represents is the \textit{rate} of local environment changes over time for the \textit{i}-th SOAP center. This allows highlighting the relevant dynamic phenomena occurring along the trajectory, notably discriminating between local environments characterized by a constant variation and those exhibiting an increasing/decreasing variation.

To increase the signal-to-noise ratio (S/N), both ${\lambda}_{i}(t)$ and $\dot{\lambda}_{i}(t)$ are pre-processed by employing the Savitzky-Golay\cite{press1990savitzky} filter from the SciPy\cite{2020SciPy-NMeth} python package, thus obtaining smoothed signals. A common polynomial order parameter of p = 2 is used for each signal ${\lambda}_{i}(t)$, while different time-windows are chosen depending on the analyzed system, in order to reach a compromise between an acceptable S/N value and a sufficiently smaller window compared to the length of signal (Fig. S1 for details). Chosen the time-window for ${\lambda}_{i}(t)$, to adequately smooth its first-derivative $\dot{\lambda}_{i}(t)$ we keep the same time-window and use two applications of the filter (following a general rule: for the {n}-th derivative, use at least n+1 applications of the filter).
\subsection{\label{sec:level2}Dynamics domains detection}
After increasing S/N, an ML-based analysis is performed on ${\lambda}_{i}(t)$ data to detect relevant dynamics domains in each system. As a clustering method, we opted to use the KMeans algorithm from the Scipy python package.\cite{lloyd1957least, macqueen1967classification} KMeans requires to pre-determine the number K of clusters to be created in the process. Here, with the aim of guaranteeing a wide variety of micro-clusters dynamics regardless of the analysed system, we start anyway from $K=10$ clusters. On the basis of the exchange probability matrix and the dendrogram associated to clusters interconnection, then we hierarchically merged the K clusters \textit{a posteriori}. The exchange probability matrix contains, indeed, the percentage probability of a unit \textit{i} belonging to a given cluster to persist in that cluster or to jump to another cluster in the sampling timestep $\Delta t$; from this, by means of an "average" linkage method, we built the associated dendrogram connecting the dynamics domains which have a high probability of exchanging elements. Ultimately, to establish the cut-off point of the dendrogram, we used the Elbow Curve Method as an indicative guideline for selecting the optimal number of clusters $k$ (see Supporting Fig. S2). However, for completeness, all the steps leading from the starting $K = 10$ clusters to the final $k$ clusters are reported in Supporting Fig. S3 and Fig. S4.

On the other hand, the domain recognition on $\dot{\lambda}_{i}(t)$ data has been performed \textit{via} a different approach. Obtained the $\dot{\lambda}_{i}$ distribution and the associated Kernel Density Estimate (KDE) for each system, we divide the KDE in deciles and consider the first and the tenth deciles to detect units significantly falling far from the mean local environment variation rate. The first decile and the tenth decile capture units highly decreasing and increasing, respectively, their local environment changes. This provides a clear distinction between domains moving toward more dynamic and those moving toward less dynamic configurations. 
\subsection{\label{sec:level3}Molecular dynamics (MD) simulations}
We test our \textit{$\tau$}SOAP analysis on MD trajectories obtained for different systems with non-trivial and different dynamics: \textit{i.e.}, a water-ice interface system in correspondence of the transition temperature, a gold nanoparticle at 200 K, a copper surface at 700 K, and DPPC lipid bilayer where liquid and gel phases coexist at 293 K of temperature. 

The atomistic ice/liquid water phase coexistence at the solid/liquid transition temperature is obtained by employing the direct coexistence technique\cite{ladd1977triple, ladd1978interfacial} using the GROMACS software\cite{hess2008gromacs}. In order to model both the ice and the liquid water phase, the TIP4P/ICE water model\cite{abascal2005potential} is used. The direct coexistence technique is based on the idea to put in contact two or more phases (in this case, the phase of ice $I_\textit{h}$ and the liquid water phase) in the same simulation box and at constant pressure. Since the energy is constant at $T = 268$ K, while the system melts at $T = 269$ K\cite{garcia2006melting}, we set the temperature at $T = 268$ K and keep it constant by means of the v-rescale thermostat with a relaxation time of $t = 0.2$ ps.\\
To get the initial configuration of ice $I_\textit{h}$ the \textit{Genice} tool proposed by Matsumoto \textit{et al.}\cite{matsumoto2018genice} is used, which generates a hydrogen-disordered lattice with zero net polarization satisfying the Bernal-Fowler rules. The solid lattice is equilibrated by performing a 10 ns-long anisotropic \textit{NPT} simulation at ambient pressure ($1$ atm). The c-rescale barostat\cite{bernetti2020pressure} is used with a time constant of $t = 20$ ps and compressibility of $9.1*10^{-6}$ $bar^{-1}$. On the other hand, the liquid phase is obtained from the same initial configuration of ice $I_\textit{h}$, but performing a \textit{NVT} simulation at $T = 400$ K in order to quickly melt the ice slab. Then, a 10 ns long simulation at $T = 268$ K is performed to equilibrate the liquid phase, using the c-rescale barostat in semi-isotropic conditions and compressibility of $4.5*10^{-5}$ $bar^{-1}$. Since the initial ice slab is composed of 1024 water molecules, both the solid and liquid phases have the same number of molecules and box dimensions. The two phases are put in contact and, then, the system is equilibrated for $t = 10$ ns employing the c-rescale pressure coupling at ambient pressure with the water compressibility ($4.5*10^{-5}$ $bar^{-1}$). The production \textit{NPT} is carried out in semi-isotropic conditions, applying the pressure only in the direction perpendicular to the ice/water interface, thus reproducing the strictly correct ensemble for the liquid-solid equilibrium simulation by the direct coexistence technique.\cite{frenkel2013simulations} Finally, a 100 ns-long production run is performed, with a sampling time interval of $\Delta t = 0.1$ ns. 

The second case study analyzed in this work is an icosahedral Gold nanoparticle (Au-NP) composed of 309 atoms. The parameterization of the model is performed according to the Gupta potential\cite{gupta1981lattice}. The Au-NP system is simulated for $t = 2$ $\mu$s at $T = 200$ K sampling every $\Delta t = 0.1$ ns using the LAMMPS software\cite{thompson2022lammps}. The details are described in reference\cite{rapetti2022}.

The third system, the atomistic model of Copper FCC surface Cu(210), is composed of 2304 Cu atoms and simulated at $T = 700$ K. A Neural Network potential built by means of the DeepMD platform\cite{wang2018deepmd} is employed to perform Deep-potential MD simulations of the Cu(210) surface with the LAMMPS software\cite{thompson2022lammps}, as reported in reference\cite{cioni2022}. The MD trajectory is conducted for $150$ ns, using a sampling time interval of $\Delta t = 0.3$ ns. 

Finally, the last case study is a DPPC lipid bilayer composed of 1152 lipids simulated at $T = 293$ K. As detailed in reference\cite{capelli2021data}, DPPC lipids are simulated and parametrized in explicit water by using the Martini2.2\cite{marrink2007martini} Coarse-Grained (CG) force field. The CG-MD simulation is performed for $t = 1$ $\mu$s and sampled every $0.1$ ns with the GROMACS software\cite{hess2008gromacs}. However, in our analysis, we use the last $500$ ns of MD trajectory.     

\section{Results and Discussion}
Herein, we use the descriptor \textit{$\tau$}SOAP to elucidate the dynamics of atomic/molecular structural environments which are often key determinants in global materials performances. In order to show the whole picture of dynamic information that can be extracted from \textit{$\tau$}SOAP signals, we analyze MD trajectories of different systems exhibiting various structures and non-trivial behaviors, thus indicating the transferability of such approach to a wide range of materials. In particular, we first focus on ice/liquid water coexistence at the solid/liquid transition temperature, where structural and dynamic properties continuously alternate from solid-like to liquid-like character\cite{bryk2002ice}. We also carry out our analysis on systems revealing solid-like dynamics, such as metal nanoparticles and surfaces well below the melting point. Ultimately, a fluid-like soft system is included by testing our approach on a lipid bilayer below the gel-to-liquid transition temperature. 

\subsection{\label{sec:level2}Into the Dynamics of Ice/Liquid Water Phase Coexistence \textit{via} \textit{$\tau$}SOAP Signal} 

\begin{figure*}[ht!]
\centering
\includegraphics[width=18cm]{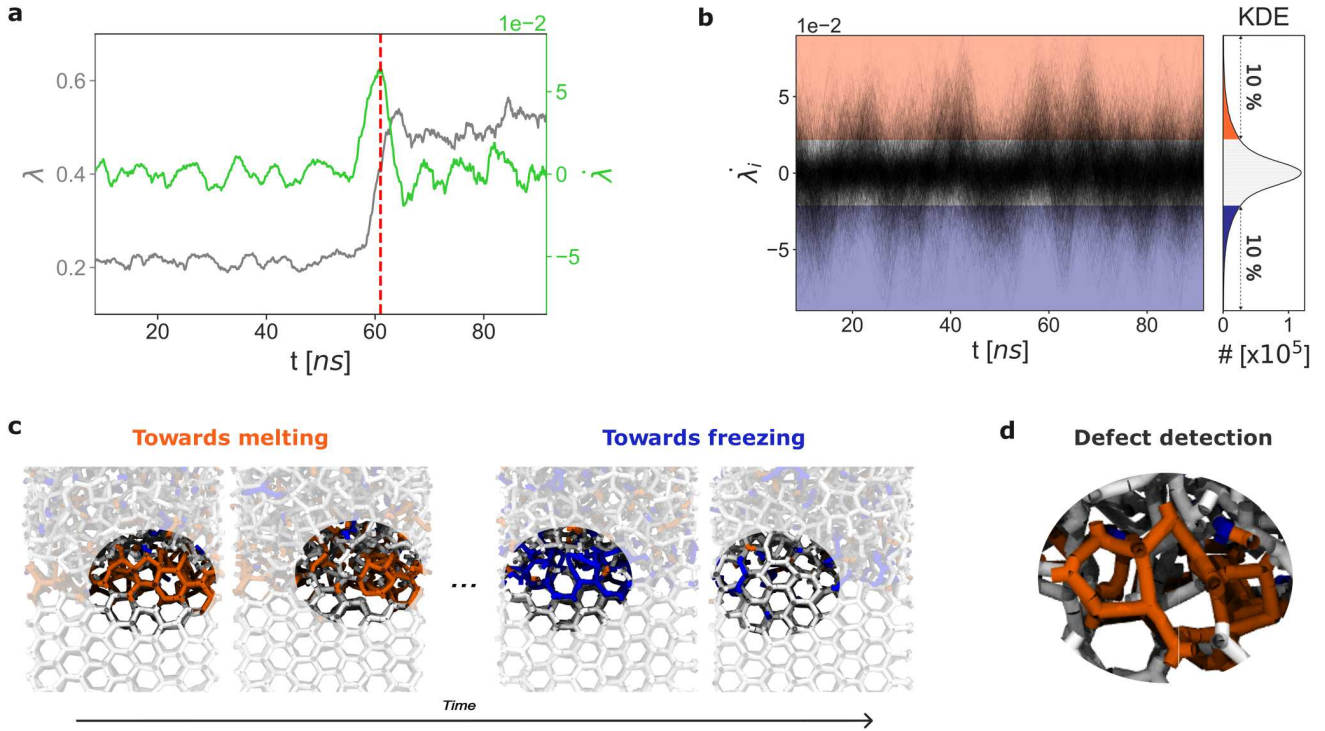}
\caption{First time-derivative of \textit{$\tau$}SOAP signal for ice/liquid water system. (a) \textit{$\tau$}SOAP (${\lambda}(t)$) and its first time-derivative ($\dot{\lambda}(t)$) profiles associated with the same oxygen atom are shown in gray and green, respectively. (b) $\dot{\lambda_{i}}(t)$ signals and their KDE estimated for all the oxygen atoms in Fig. \ref{fig:fig01}a. Clustering color code: (i) blue for environments corresponding to the first decile; (ii) orange for those corresponding to the tenth decile; (iii) white for $\dot{\lambda_{i}}$ values in all the other deciles. (c) MD snapshots displaying blue, orange, and white domains. Left: local detail of the orange cluster evolving toward melting (first and second snapshots). Right: local detail of blue cluster associated with a small disordered region evolving toward freezing (third and fourth snapshots). (d) Orange local environments identify ice molecules moving out of hexagonal ice configurations.}
\label{fig:fig02}
\end{figure*}

We start testing \textit{$\tau$}SOAP on a system where crystalline ice and liquid water coexist at the solid/liquid transition temperature, while exhibiting a dynamic equilibrium between solid-like and liquid-like regime. We analyze a simulation box, in periodic boundary conditions, having 1024 hexagonal ice (\textit{Ih}) molecules in contact with 1024 liquid water molecules (see Fig.~\ref{fig:fig01}a) at $T = 268$ K. We consider 1001 consecutive frames sampled every $\Delta t = 0.1$ ns along 100 ns of an MD trajectory. As a first step, we compute the SOAP vectors for the oxygen atoms of each water molecule (2048 TIP4P/ICE water molecules) along all frames of the trajectory (see Methods section for details). 
\textit{$\tau$}SOAP signals are then estimated by capturing the variations of local SOAP environments in $\Delta t = 0.1$ ns (see Eq. \ref{eq:six}). Fig.~\ref{fig:fig01}c reports, on the left, the resulting ${\lambda}_{i}(t)$ time-profiles related to each of the 2048 oxygen atoms, while, on the right, it shows the ice/liquid water MD snapshots at $t = 59$ ns,  $t = 63$ ns, and  $t = 65$ ns. Notably, three distinct ${\lambda}_{i}(t)$ profiles are highlighted in Fig.~\ref{fig:fig01}c, left: (i) the black signal oscillating around ${\lambda}_{i} = 0.2$; (ii) the cyan signal laying on the highest ${\lambda}_{i}$ region; and (iii) the crimson signal which rapidly passes, at $t \sim 60$ ns, from low to high ${\lambda}_{i}$ values. The oxygen atoms related to the latter three ${\lambda}_{i}(t)$ profiles are instead depicted on the MD snapshots in Fig.~\ref{fig:fig01}c, right, with the respective color code, i.e., black, cyan, and crimson. The visualization of these selected atoms clearly shows that the black and cyan oxygen units belong to the ice and liquid water phase, respectively, regardless of the displayed time steps. On the other hand, the identified crimson oxygen represents an atom involved in the ice/liquid water transition occurring at $t \sim 60$ ns. By lightening the behavior of such atoms, we attempt to emphasize the potential understanding provided by \textit{$\tau$}SOAP descriptor on the single unit dynamics: following the time variation of atomic structural environments, \textit{$\tau$}SOAP allows both to distinguish atoms belonging to different phase states and to capture those one undergoing phase transitions.

In order to systematically detect the complete scenario of distinct dynamics behaviors in our water system, an ML-based analysis is carried out on all \textit{$\tau$}SOAP signals. The results of the clustering investigation, performed \textit{via} the KMeans algorithm, are shown in Fig.~\ref{fig:fig01}d. The final four identified clusters (gray, crimson blue, and cyan) are displayed both on the time-series of the ${\lambda}_{i}(t)$ data (Fig.~\ref{fig:fig01}d: left) and on the ${\lambda}_{i}(t)$ data distribution reported with the correlated KDE (right). As already pointed out, such four different dynamic domains identify those water molecules undergoing specific transitions: \textit{i.e.}, instantaneously changing their local structural environments. In particular, the gray domain is dominated by oxygen units that are characterized  by low ${\lambda}_{i}$ values along the complete trajectory, i.e. by a weak variability of their local atomic environments. On the other hand, oxygen atoms showcasing high changes of their structural atomic distributions, and hence high values of ${\lambda}_{i}$, belong to the blue cluster or cyan domain. Oxygen units that, instead, tend to reveal medium values of ${\lambda}_{i}$ - because of their transition from one $\lambda_{i}$ regime to the other one - are classified into a distinct crimson cluster. Ultimately, the cyan domain  is detected as a different cluster of units having higher local environment changes. The graphical representation of such clusters is shown in Fig.~\ref{fig:fig01}f through an MD snapshot. Not surprisingly, the gray cluster, characterized by the lowest \textit{$\tau$}SOAP signal, corresponds to the ice phase; the blue domain is mainly correlated to the liquid phase; and the crimson one, including oxygen atoms with 0.2 $ < {\lambda}_{i} < $ 0.4, is instead located at the solid/liquid interface. Finally, the cyan cluster, although sited in the same region of the liquid phase (blue cluster), is identified as presenting a different dynamic behavior. In the considered ice/liquid water system, the exchange probabilities among the final four clusters are displayed in the matrix in Fig.~\ref{fig:fig01}e: although the oxygen atoms exhibit a probability higher than $94\%$ to remain in the belonging cluster (probabilities on the matrix diagonal), no negligible transient events occur between red-blue and cyan-blue clusters, demonstrating that a percentage of oxygen population is involved into instantaneous transitions among dynamics domains (out of diagonal probabilities). 

After detecting the main dynamics clusters based on \textit{$\tau$}SOAP signals, ${\lambda}_{i}(t)$, we carry out a further domain recognition analysis based on $\dot{\lambda}_{i}(t)$, that is, the instantaneous \textit{rate} of local environment variations ${\lambda}_{i}(t)$. The key information which can be gathered from the time-derivative of ${\lambda}_{i}(t)$ is pointed out in Fig.~\ref{fig:fig02}a, where an explicative example is reported. Here, both ${\lambda}_{i}(t)$ and $\dot{\lambda}_{i}(t)$ time-profiles are associated with the same oxygen atom \textit{i}: in gray, ${\lambda}_{i}(t)$ shows the atom undergoing the phase transition at $t \sim 60$ ns when the time-signal significantly and rapidly passes from the low to the high ${\lambda}_{i}$ value region; in green, the first time-derivative of the gray profile exhibits a peak in correspondence of the phase transition, while fluctuating around zero in both the initial and the final stages of the trajectory. Clearly, $\dot{\lambda}_{i}(t)$ tracks a notable signal leading up to a substantial dynamic change in the system. The first time-derivative, indeed, offers a neat discrimination between small oscillations of ${\lambda}_{i}(t)$ - which are intrinsic to the constituent units, independently from the proper dynamics domain - and large fluctuations driving significant changes in the atomic structure. Notably, $\dot{\lambda}_{i}(t)$ also provides a detailed comprehension of the \textit{directionality} of the local environment variations, \textit{i.e.}, on the evolution of the material structures. While the presence of a peak, \textit{i.e}, of a large fluctuation in $\dot{\lambda}_{i}(t)$ profile, suggests that a relevant event is occurring in that time interval, the sign of such fluctuation points out the evolution of a structural environment: a positive sign indicates that the atom is undergoing a local re-configuration toward a more dynamic domain; a negative sign means that a local environment re-configuration toward a more static domain is occurring.

Fig.~\ref{fig:fig02}b shows, on the left, the time-profiles of  $\dot{\lambda}_{i}(t)$ related to each of the 2048 oxygen atoms, while, on the right, the KDE of the $\dot{\lambda}_{i}$ data distribution. We color in blue and orange the domains corresponding to the first and the tenth decile, respectively, while we merge all the other deciles in a single white cluster. It is worth noting that the KDE distribution has a peak approximately in correspondence of $\dot{\lambda}_{i} = 0$, indicating that the local environment variations -${\lambda}_{i}(t)$- are, on average, constant. On the other hand, atoms that significantly increase or decrease, frame by frame, their local environment changes are captured by positive (in the orange region) or negative peaks (in the blue region), respectively. In Fig.~\ref{fig:fig02}c, we visualize these three different domains (blue, orange and white) on some snapshots along the MD trajectory, thereby showing that the positive and negative peaks allow characterizing melting and freezing phenomena occurring within small solid-like and liquid-like regions. In the first snapshot of Fig.~\ref{fig:fig02}c, we represent a small portion of the orange cluster at the ice/liquid water interface. Such orange oxygen solid-like atoms are exhibiting positive peaks, \textit{i.e.}, they are undergoing rearrangements toward more dynamic configurations. Accordingly, in the second snapshot those rings appear as broken, thus proving a melting-type process. In the third snapshot of Fig.~\ref{fig:fig02}c,  we report, instead, an example of oxygen units associated with the blue cluster (negative peaks). Here, the small blue domain displayed at the solid-liquid interface should include those atoms that are evolving toward more static configurations. Indeed, as shown in the fourth snapshot, an ordered ring structure forms, thus  reproducing a typical freezing phenomenon. 
Ultimately, Fig.~\ref{fig:fig02}d shows a further detail potentially revealed by our analysis. In particular, water molecules exhibiting a high positive rate of change of their local SOAP environment (high $\dot{\lambda}_{i}$) turned out to be also associated with ice molecules that, at the interface with liquid water, undergo transitions out of the typical hexagonal packing: \textit{i.e.}, forming interface ice defects (Fig.~\ref{fig:fig02}d)\cite{moritz2021microscopic}. In summary, besides capturing the local atomic re-arrangements and characterizing their evolution, $\dot{\lambda}_{i}(t)$ seems to be also promising for defects detection purposes. 

The previous results suggest how \textit{$\tau$}SOAP descriptor and its first time-derivative are possible strategies to unveil some microscopic phenomena occurring at the ice/water interface in a dynamic equilibrium. In particular, by reliably detecting local fluctuations along with rearrangements and their evolution, the time-variations of structural atomic environments show considerable potential for tracking crystallization or melting processes from MD trajectories.   

\subsection{\label{sec:level3}Application to discrete solid-like dynamics}
\begin{figure*}[ht!]
\centering
\includegraphics[width=18cm]{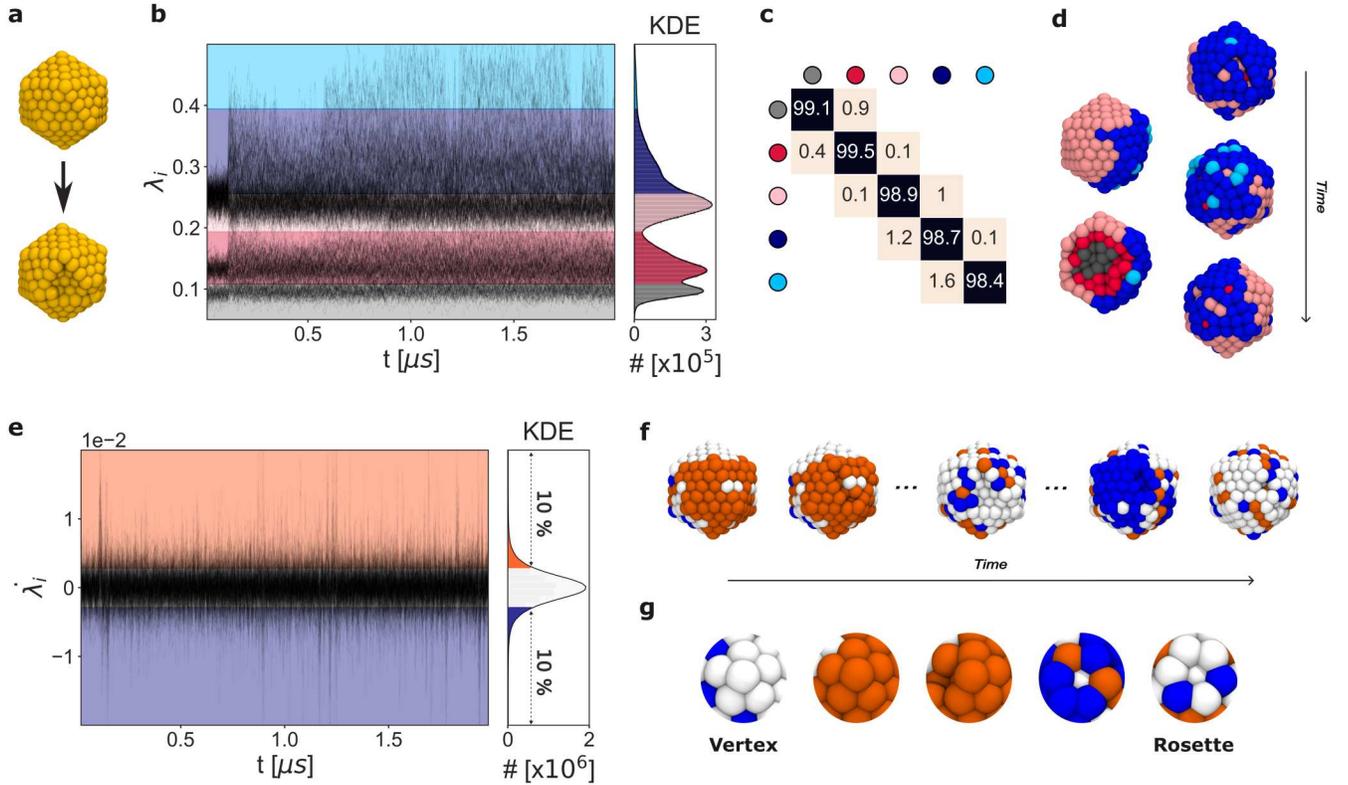}
\caption{\textit{$\tau$} SOAP-based analysis on 309-atoms icosahedral Gold nanoparticle (Au-NP). (a) MD snapshots of ideal Au-NP (top), and equilibrated one at $T = 200$ K (bottom). (b) ${\lambda}_{i}(t)$ profiles and the related KDE for the Au atoms in the system (a). The final $k = 5$ macro-clusters identified by KMeans are shown in gray, crimson, pink, blue, and cyan. (c) Exchange probability matrix of the final $k = 5$ detected macro-clusters. (d) MD snapshots with the five main clusters identified in (b): inner core in gray, interface region between the inner core and the outermost layer in crimson, more static surface face in pink, more dynamic surface face in blue, atoms undergoing the highest local environments changes in cyan. (e) Domain detection based on $\dot{\lambda}(t)$ profiles and their KDE: blue domain is associated with the first decile, the orange domain is linked to the tenth decile, and the white domain includes $\dot{\lambda}(t)$ in all the other deciles. (f) MD snapshots displaying the emergence of blue, orange, and white domains along the MD trajectory. On the left, the predominance of the orange cluster before (first snapshot) and during the symmetry breakdown (second snapshot). The central snapshot exhibits a prevalence of white domain, together with a balance between orange e blue ones. On the right, a prevalence of the blue domain can be observed (fourth snapshot) before the formation of a more static configuration (white cluster: fifth snapshot). (g) Blue, orange, and white domains associated with the rearrangement, over time, of a local configuration from "vertex" to "rosette".}
\label{fig:fig03}
\end{figure*}

\begin{figure*}[ht!]
\centering
\includegraphics[width=18cm]{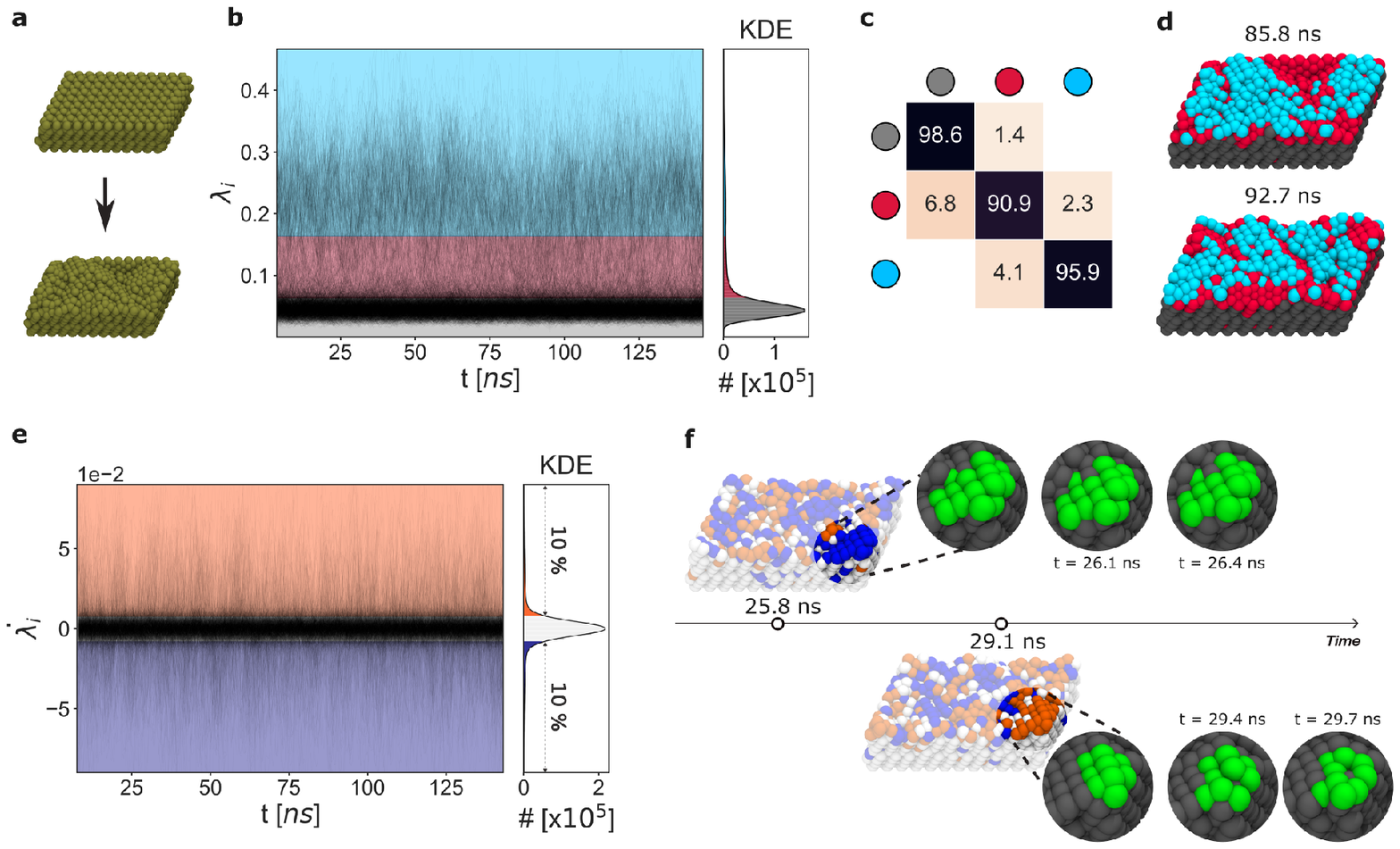}
\caption{\textit{$\tau$}SOAP based analysis on a Copper FCC surface, Cu(210), composed of 2304 atoms. (a) MD snapshots of ideal Cu(210) surface (top), and equilibrated one at $T = 700$ K (bottom). 
(b) ${\lambda}_{i}(t)$ signals and the related KDE for Cu atoms of system in (a). The final $k = 3$ macro-clusters identified by KMeans are shown in gray, crimson, and cyan. 
(c) Exchange probability matrix of the final $k = 3$ detected macro-clusters. 
(d) MD snapshots showing the three main clusters identified in b): crystalline bulk in gray, sub-surface region in crimson, more dynamic surface atoms in cyan. 
(e) Domain detection based on $\dot{\lambda}(t)$ profiles and the related KDE: the blue domain is associated to the first decile, the orange cluster is linked to the tenth decile, the white domain is related to all the other deciles. 
(f) MD snapshots of blue, orange and white domains (in transparency) related to two different frames. Top: at $t = 25.8$ ns, the circled portion of the surface exhibits a prevalence of blue domain, thus predicting stable reconfigurations in the two successive frames (green atoms). 
Bottom: at $t = 29.1$ ns, the same circled portion exhibits a predominance of orange cluster, thus predicting dyanmic reconfigurations in the two successive frames (green atoms).}
\label{fig:fig04}
\end{figure*}

As completely different test cases, we test our approach on systems revealing solid-like dynamics. We discuss the results of our analysis applied on MD trajectories of (i) a 309-atoms icosahedral Gold nanoparticle, denoted as Au-NP, at 200 K (Fig. \ref{fig:fig03}a), and (ii) a Copper Cu(210) FCC surface at 700 K of temperature (Fig.~\ref{fig:fig04}a).  

Regarding the case (i), we analyse 20000 consecutive frames of a 2-$\mu$s long MD trajectory sampled every $\Delta t = 0.1$ ns at $T = 200$ K. It is well known that metal nanoparticles may exhibit a not-trivial dynamics at room and at even sufficiently lower temperatures. Although the reduced atomic motion and, accordingly, the more stabilized ideal icosahedron architecture, some local fluctuations and atomic rearrangements can be observed in a Au-NP even at $T = 200$ K. \textit{$\tau$}SOAP signals in Fig.\ref{fig:fig03}b, indeed, present a sudden increase after $\sim 0.1$ $\mu$s, demonstrating that some atoms are experiencing intense instantaneous local environment variations. Our cluster analysis on $\lambda_{i}(t)$ recognises five main dynamics domains whose transition probabilities are reported in Fig. \ref{fig:fig03}c. This transition matrix proves a negligible attitude of the gold atoms to transfer from/toward diverse dynamics domains, while preferring to remain in their own cluster with probabilities higher than 98.4 $\%$. The MD snapshots in Fig. \ref{fig:fig03}d show that these clusters well identify distinct dynamics behaviors and structural domains. First, the cluster analysis is able to accurately distinguish the inner core of the Au-NP (in gray), namely a more static region characterised - not surprisingly - by low $\lambda_{i}(t)$ values along the whole simulation, from an interface region (in crimson) between the inner core and the outermost layer. Second, such clustering approach sharply separates the surface of the Au-NP in two coexisting regions (pink and blue) related to different characteristic $\lambda_{i}(t)$. While the pink face turned out to be more static, the blue domain reliably detects the portion of the surface where a fracture formation may occur, breaking down the symmetry (Fig. \ref{fig:fig03}d, second MD snapshots on the right). Interestingly, \textit{$\tau$}SOAP also identifies some local events such as the formation of concave "rosettes" (a vertex, having five neighbors in an ideal icosahedron, penetrates inside the NP surface, thus passing to six neighbors). In Fig. \ref{fig:fig03}d (third snapshot on the right), two rosettes can be observed as belonging to a more dynamic cluster - highly varying their local environments - (in blue), while the associated vertices are identified as more stable (in crimson).

Furthermore, the estimation of $\dot{\lambda}_{i}$ (Fig. \ref{fig:fig03}e) provides interesting details on the dynamic evolution of the system. A quite large percentage of Au atoms is characterized by a constant variations of their surrounding environment ($\lambda_{i}(t)$ = cost, and $\dot{\lambda}_{i} \sim 0$). Rare and sharp fluctuations are anyhow remarkable. To qualitatively illustrate some of these $\dot{\lambda}$ peaks, five MD snapshots presenting different predominant domains are shown Fig. \ref{fig:fig03}f. In the first and second snapshots, the orange domain prevails, suggesting that the atoms belonging to that cluster are collectively involved in a significant increase of the instantaneous local environment variations (positive $\dot{\lambda}_{i}$). Indeed, this predicts the symmetry breaking shown in the second snapshot. However, the prevalence of the white domain,  i.e. $\dot{\lambda}_{i} \sim 0$, along with a balance between positive (orange) and negative peaks (blue), establishes a dynamic equilibrium leading to no relevant events along several trajectory frames (one example is presented in the third snapshot). In the last two snapshots, instead, the prevalence of the blue domain (fourth snapshot) indicates a significant collective decrease of the instantaneous local variations (negative $\dot{\lambda}_{i}$), thus predicting the evolution of the related atoms toward more static environments, as shown in the final snapshot. Ultimately, the information on the directionality of local rearrangements is also highlighted in Fig. \ref{fig:fig03}g: while the orange cluster (positive $\dot{\lambda}_{i}$ values) includes a vertex evolving toward a less stable configuration where a missing atom appears, the blue domain (negative $\dot{\lambda}_{i}$) predicts the rearrangement of the structure toward a stable rosette-like configuration.

For case (ii), we use 502 consecutive frames of 150 ns long MD simulation of a Cu(210) surface composed of 2304 Cu atoms (Fig.~\ref{fig:fig04}a) sampled every $\Delta t = 0.3$ ns at $T = 700$ K.  Although metals tend to be traditionally considered as hard matter, it is now well known that their constituent surface atoms may exhibit a non-trivial dynamics, undergoing rearrangements well below the melting temperature. \cite{cioni2022,daff2009computer} Our clustering procedure applied on \textit{$\tau$}SOAP profiles identifies three main domains related to Cu atoms exhibiting very competing behaviors (Fig.~\ref{fig:fig04}b): one dense and more static cluster in gray along with two less populated but more dynamic domains in red and cyan. The exchange probability matrix in Fig.~\ref{fig:fig04}c points out that the transient events among diverse domains mainly engage Cu atoms belonging to the red and cyan clusters. Fig.~\ref{fig:fig04}d graphically represents the identified clusters at two explicative time steps, $t = 85.8$ ns and $t = 92.7$ ns: not surprisingly, the gray domain corresponds to the crystalline bulk of the Cu(210) surface, reasonably detected by our analysis as the most static with small local environment variations (low $\lambda_{i}(t)$ values); on the other hand, the surface atoms are identified as more dynamic clusters, thereby including all $\lambda_{i}(t)$ > 0.07. However, two sub-surface regions are recognized by KMeans: in crimson, a domain characterized by $ 0.07 < \lambda_{i}(t) < 0.16$, and in cyan, a cluster with the highest local environment variations. The two MD snapshots in Fig.~\ref{fig:fig04}d show the significant correlation between the crimson dynamics domain and more stable surface atomic arrangements with increased coordination. 

\begin{figure*}[ht!]
\centering
\includegraphics[width=18cm]{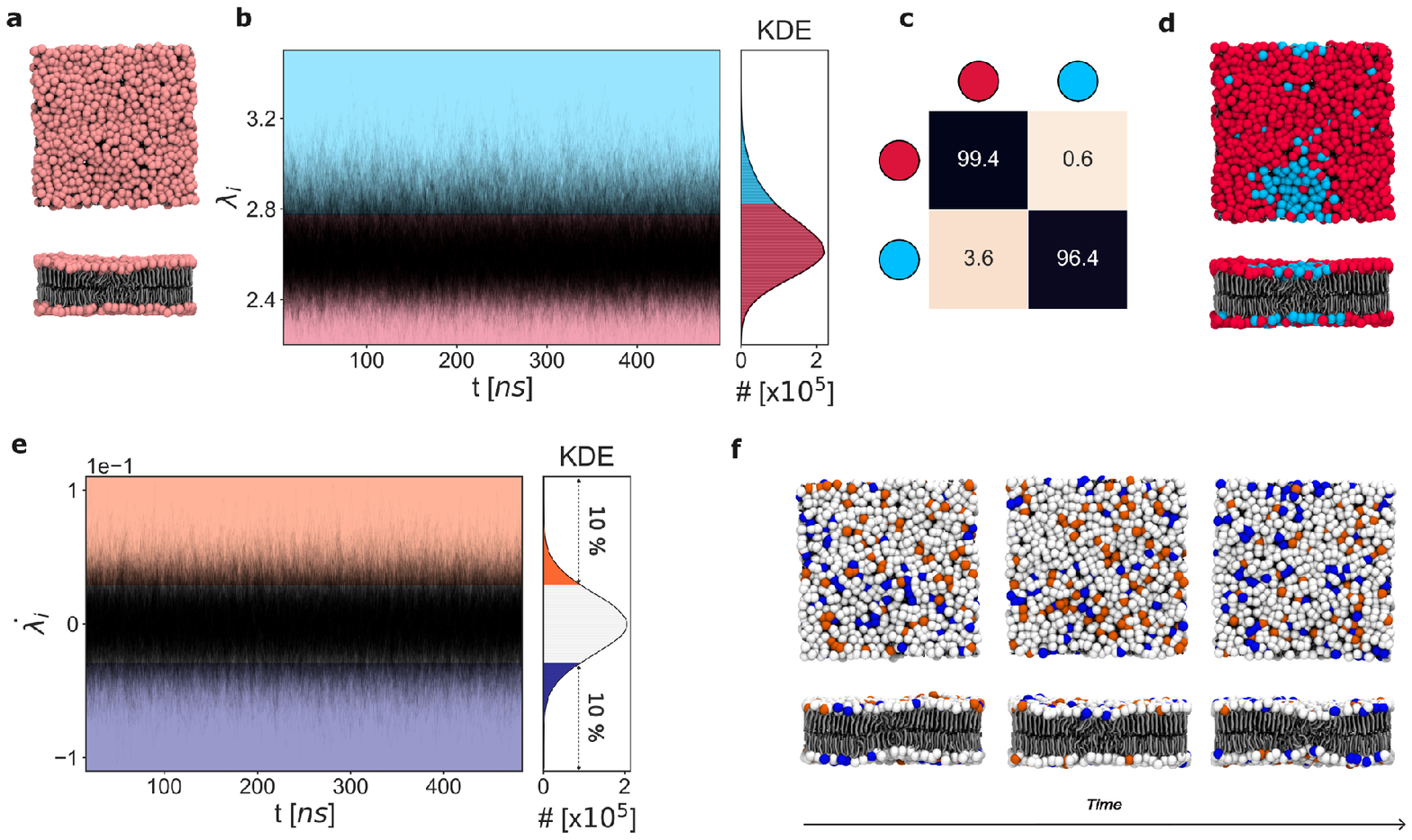}
\caption{\textit{$\tau$} SOAP-based analysis on a lipid bilayer composed of 1152 DPPC lipids at $T = 293$ K. (a) MD snapshots of DPPC lipid bilayer (top and lateral views). (b) ${\lambda}_{i}(t)$ signals and the related KDE for all the phosphate atoms of all lipid molecules in the DPPC bilayer (a), along the last $500$ ns of the MD trajectory. KMeans clustering identifies $k = 2$ final macro-clusters shown with crimson and cyan. (c) Exchange probability matrix of the final $k = 2$ macro-clusters. (d) MD snapshot of the two detected domains (same color code as (b)): gel phase in crimson, and liquid phase in cyan. (e) Domain detection based on $\dot{\lambda}(t)$ profiles and the related KDE: the blue domain is associated to the first decile; the orange cluster is linked to the tenth decile; the white domain is related to all the other deciles. (f) MD snapshots of blue, orange, and white domains related to different frames along the trajectory.}
\label{fig:fig05}
\end{figure*}

The Cu(210) domain characterization based on $\dot{\lambda}_{i}$ confirms the effectiveness of this analysis in providing some key information on the time evolution of the material structure. Fig.~\ref{fig:fig04}e highlights, also in this case, that the average rate of the local environment variations is null, \textit{i.e.}, most of the Cu atoms in Cu(210) shows a steady-state behavior of $\lambda_{i}(t)$. In addition, the cluster representation Fig.~\ref{fig:fig04}f suggests that the white domain ($\dot{\lambda}_{i} \sim 0$) is mainly correlated to the crystalline bulk. On the other hand, most of the surface atoms are highly dynamic, and consequentially a balance between domains with positive (orange) or negative (blue) $\dot{\lambda}_{i}$ is established over time. Consistent with the test cases discussed above, this dynamic balance indicates that any substantial reconfiguration toward more stable/dynamic arrangements is not occurring. Nevertheless, some cluster details are interesting to be noticed in Fig.~\ref{fig:fig04}f: the snapshot on the top, corresponding to $t = 25.8$ ns, exhibits a portion of the surface with a clear predominance of the blue domain; the zoom onto that portion clarifies its stability in the two successive sampled times ($t = 26.1$ and $t = 26.4$). On the contrary, when the same surface region is characterized, after some frames, by a prevalence of the orange domain (snapshot on the bottom at $t = 29.1$ ns), the associated atoms experience an evident rearrangement as highlighted by the green atoms onto the zoom. Again on this system, $\dot{\lambda}_{i}$ was revealed to be useful in predicting the evolution toward more static/more dynamic configurations.     

\subsection{\label{sec:level4}Phases coexistence in soft dynamical systems}
As a final test case, we apply our \textit{$\tau$}SOAP-based dynamics domain recognition on a soft system characterized by a two-phase coexistence, \textit{i.e.}, gel and liquid. Specifically, we analyze the last 500 ns of 1 $\mu$s-long CG-MD simulation of a DPPC lipid bilayer composed of 1152 self-assembled DPPC lipids (see Fig.~\ref{fig:fig05}a) at $T = 293$ K, thus considering the last 5001 consecutive frames ($\Delta t = 0.1$ ns). Although the gel-to-liquid transition temperature of a DPPC membrane is at $\sim 315$ K, here we investigate the dynamics of the lipid bilayer at a slightly lower temperature, thereby avoiding addressing the critical dynamics issues occurring at the transition temperature. 

Our clustering analysis, displayed in Fig.~\ref{fig:fig05}b on $\lambda_{i}(t)$ profiles and on the related KDE, identifies two main dynamics domains: one colored in crimson including  $\lambda_{i} < 2.8 $;  and the other one in cyan, containing the highest values of \textit{$\tau$}SOAP fingerprints. Fig.~\ref{fig:fig05}c reports, instead, the interconversion matrix between the two clusters. Beyond a small probability (~3.6 $\%$) to transient from cyan to red cluster, the lipids manifest relatively high stability to preserve, along the complete trajectory, a specific local environment variation ($\lambda_{i}$), typical of their belonging cluster. The graphical representation of the lipid bilayer in Fig.~\ref{fig:fig05}d suggests a close link between the dynamics domains and the phase states: the crimson cluster characterized by small $\lambda_{i}$ is indeed associated to a more static - gel - phase; while the cyan domain, with higher local environment variations, is connected to a more dynamic - liquid - phase. 

The further analysis on $\dot{\lambda}_{i}(t)$, reported in Fig.~\ref{fig:fig05}e, detects a predominance of the white domain along with a balance between the orange and the blue clusters over time. We recall indeed that a null time-derivative of $\lambda_{i}(t)$ represents the behavior of those units exhibiting a constant variation of their local environment, with some statistical oscillations classified in the orange and blue domain. Within such a resulting scenario, the proposed analysis predicts gel-liquid phases coexist in a dynamic equilibrium, as shown in the MD snapshots in Fig.~\ref{fig:fig05}f. In other words, the lipids are not evolving toward a more static/dynamic configuration, whereas each remains in its proper dynamics domain. 

\section{Conclusions}
Understanding the dynamics of individual units in many atomic/molecular systems is essential to understand the behavior of complex molecular systems, their physical and chemical properties, collective transitions, as well as to designing next-generation materials and molecular systems with desirable dynamical behaviors\cite{xie2019graph}. However, because of the complexity of local structural environments along with their dynamics in such systems, a general approach is still lacking. Although faithful representations of atomic neighborhood environments - such as the SOAP descriptor - are available and widely employed, here we want to draw attention to the time evolution of these structures, which is typically overlooked in molecular motif recognition procedures. 

In this work we propose an alternative perspective allowing us to track the dynamical changes in atomic structural environments of the interacting sub-units, thus enhancing the detection of dynamics domains and emerging phenomena. Building upon the SOAP descriptor, we implement \textit{$\tau$}SOAP, a new fingerprint that quantifies the variations of local SOAP environments surrounding each constituent unit along its MD trajectory. \textit{$\tau$}SOAP, indeed, retains the time information from the high-dimensional SOAP vectors, thereby aiming at emphasizing the importance of consequential events for reconstructing dynamics and detecting rare fluctuations. Coupled to an ML-based analysis, we demonstrate the potentiality of such an approach to identify domains with different structural and dynamical behaviors. Ranging from an ice/liquid water interface system where solid-like and fluid-like domains coexist in a dynamic equilibrium, to solid-like materials, and soft matter presenting gel and liquid coexisting phases, we prove that our analysis reliably addresses phase transitions, rare dynamic events, and coexisting phases. Moreover, by estimating the first time-derivative of \textit{$\tau$}SOAP signal, we gain further information on the direction of the local structural changes. Indeed, besides detecting local rearrangements, the first time-derivative of \textit{$\tau$}SOAP enables the characterization of their evolution toward either highly or weakly dynamic environments. However, we envisage a paradigm change in order to improve the SOAP-based data investigation for pattern recognition with temporal information and more dynamics understanding. 

Nonetheless, \textit{$\tau$}SOAP-based investigation presents some limitations. Although \textit{$\tau$}SOAP signal tracks the evolution of each constituent unit along the whole MD trajectory, thus providing time history data, the coupled ML-based approach relies on the instantaneous values of local environment changes, without performing a time-series clustering for identifying dynamics domains. Importantly, time-series clustering and classification based on the frequency/duration of local environment variations could have a striking advantage in discriminating fluctuations leading up to significant structural changes in the system. Notably, by including in our ML-based framework the first time derivative of \textit{$\tau$}SOAP we start providing some further insights on predicting the evolution of local changes, and specifically how selected environments reconstruct or evolve in time. In summary, our approach turned out to be robust and versatile to capture fluctuating environments from SOAP spectra in a variety of systems by means of a completely agnostic and data-driven analysis.  

\section*{Authors' contributions}
G.M.P. conceived this research and supervised the work. C.C. developed the descriptor and implemented the analysis. C.C., M.C. D.R., and A.C. performed the simulations and the analyses. All authors analyzed and discussed the results. C.C., A.C., and G.M.P. wrote the manuscript.

\begin{acknowledgments}
G.M.P. acknowledges the support received by the European Research Council (ERC) under the European Union’s Horizon 2020 research and innovation program (Grant Agreement no. 818776 - DYNAPOL) and by the Swiss National Science Foundation (SNSF Grant IZLIZ2$\_183336$).
\end{acknowledgments}

\section*{Data Availability Statement} 
Complete details of all molecular models used for the simulations, and of the simulation parameters (input files, etc.), as well as the complete \textit{Time}SOAP analysis code, are available at: \url{https://github.com/GMPavanLab/TimeSOAP} (this temporary folder will be replaced with a definitive Zenodo archive upon acceptance of the final version of this paper). Further details on the analyses are provided in the Supporting Information file. 

\nocite{*}
\bibliography{bibliography}

\end{document}


\title{Supporting Information for: \textit{Time}SOAP:
Tracking high-dimensional fluctuations in complex molecular systems \textit{via} time-variations of SOAP spectra}
\author{Cristina Caruso}
\affiliation{Department of Applied Science and Technology, Politecnico di
Torino, Corso Duca degli Abruzzi 24, 10129 Torino, Italy}

\author{Annalisa Cardellini}
\affiliation{Department of Innovative Technologies, University of Applied
Sciences and Arts of Southern Switzerland, Polo Universitario Lugano, Campus Est, Via la Santa 1, 6962 Lugano-Viganello, Switzerland}

\author{Martina Crippa}
\affiliation{Department of Applied Science and Technology, Politecnico di
Torino, Corso Duca degli Abruzzi 24, 10129 Torino, Italy}

\author{Daniele Rapetti}
\affiliation{Department of Applied Science and Technology, Politecnico di
Torino, Corso Duca degli Abruzzi 24, 10129 Torino, Italy}

\author{Giovanni M. Pavan}
\email{corresponding author: Giovanni M. Pavan
(giovanni.pavan@polito.it)}
\affiliation{Department of Applied Science and Technology, Politecnico di
Torino, Corso Duca degli Abruzzi 24, 10129 Torino, Italy}
\affiliation{Department of Innovative Technologies, University of Applied Sciences and Arts of Southern Switzerland, Polo Universitario Lugano, Campus Est, Via la Santa 1, 6962 Lugano-Viganello, Switzerland}

\maketitle
\begin{table}[t]
    \centering
    \begin{tabular}{|c|c|c|c|c|c|}
    \hline
        \textbf{SYSTEM} & \textbf{$r_{cut}$ [{\AA}]} &\shortstack{\textbf{SOAP} \\ \textbf{center}} & \shortstack{\textbf{MD[ns]}} & \shortstack{\textbf{\# of sampled}\\\textbf{frames} }& \shortstack{\textbf{Sampling}\\\textbf{$\Delta t$ [ns]}}  \\ 
        \hline
        \shortstack{Ice/Liquid \\ water} & 10 & OW & 100 & 1001 & 0.1 \\\hline
        \shortstack{Au-NP} & 4.48 & Au & 1000 & 20000& 0.1 \\ \hline
        \shortstack{Cu(210)} & 6 & Cu & 150 & 502 & 0.3 \\\hline
        \shortstack{DPPC \\ Lipids} & 30 & PO4 & 500 & 5001 & 0.1 \\\hline       
    \end{tabular}
    \caption{Setup details on SOAP vectors computation and MD simulation for all systems.}
    \label{tab:tab01}
\end{table}

\begin{figure*}[ht!]
\includegraphics[width=18cm]{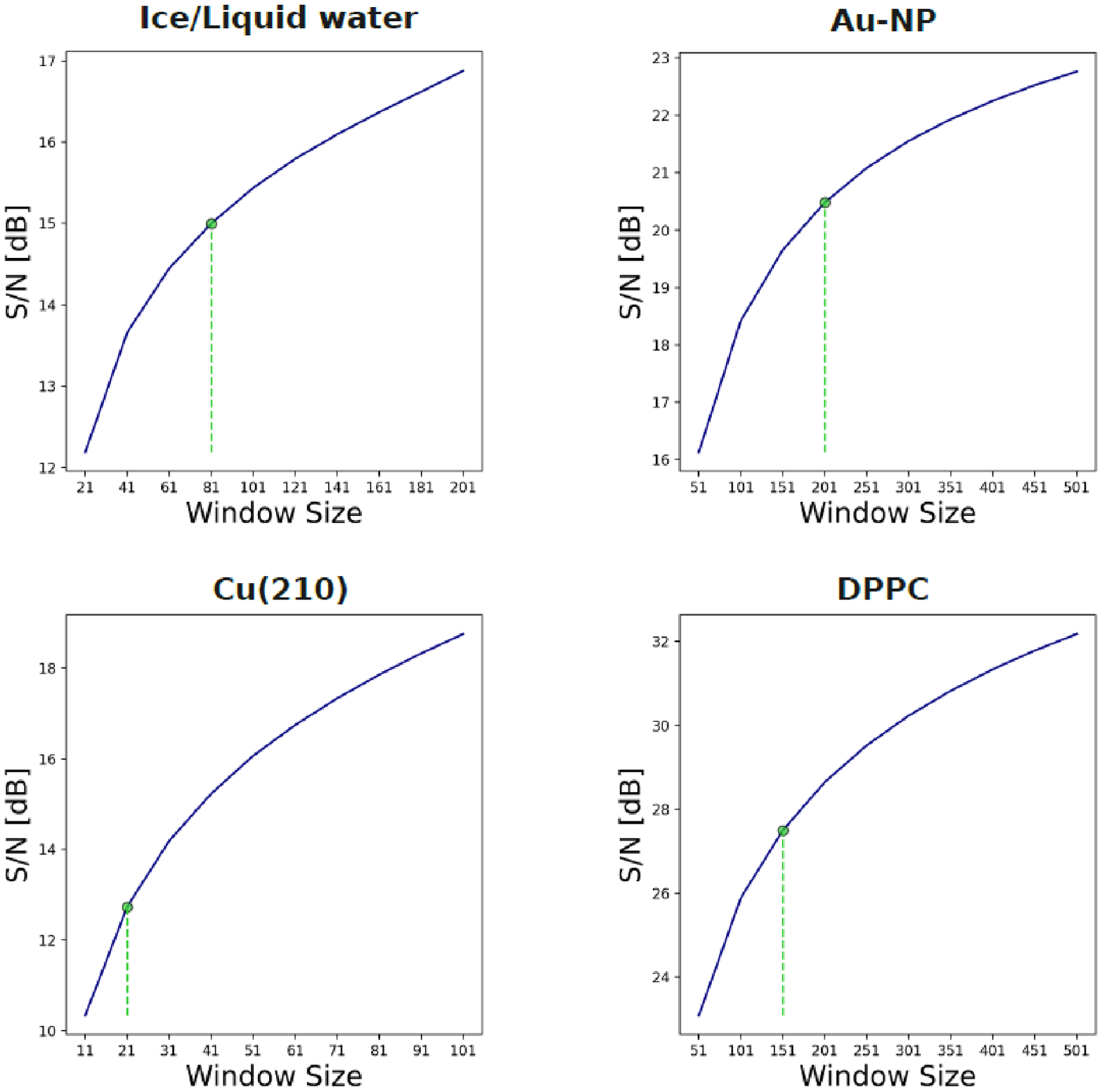}
\centering
\caption{Signal-to-noise ratio (S/R) of \textit{$\tau$}SOAP signals varying the window size of the Savitzky–Golay filter for all systems. Each value is obtained as the average of the S/R of all the \textit{$\tau$}SOAP signals filtered using that specific window. A green marker indicates the chosen window size for each system.}
\label{SIfig01}
\end{figure*}

\begin{figure*}[ht!]
\includegraphics[width=18cm]{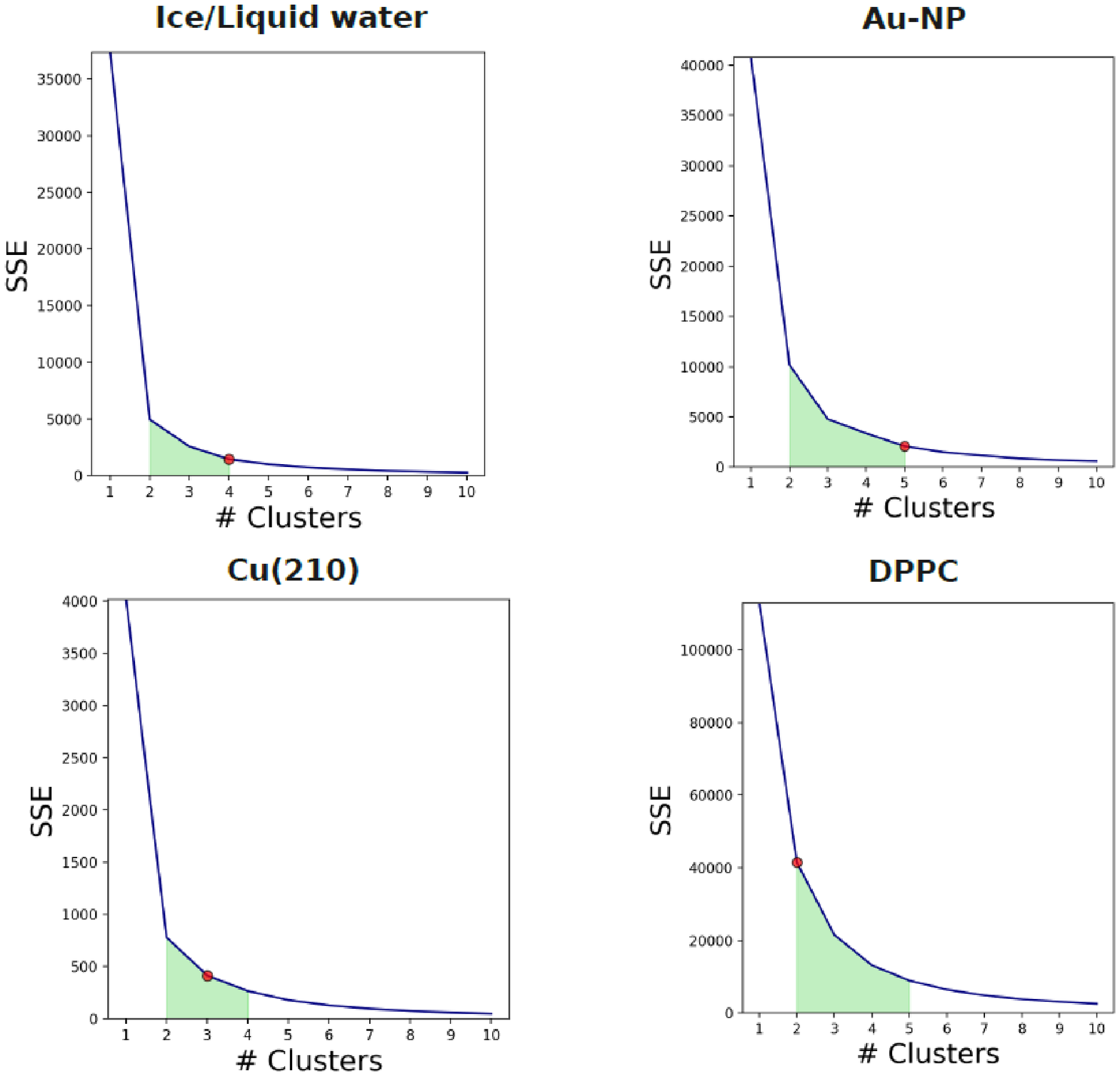}
\centering
\caption{Sum of the Squared Euclidean (SSE) distances varying the number of clusters identified by KMeans for all systems. The green area is located approximately at the elbow of the curve, showing the possible choices of the number $k$ of clusters that would optimize the clustering results according to the elbow method approach. A red marker indicates the final number of clusters we chose.}
\label{SIfig02}
\end{figure*}

\begin{figure*}[ht!]
\includegraphics[width=18cm]{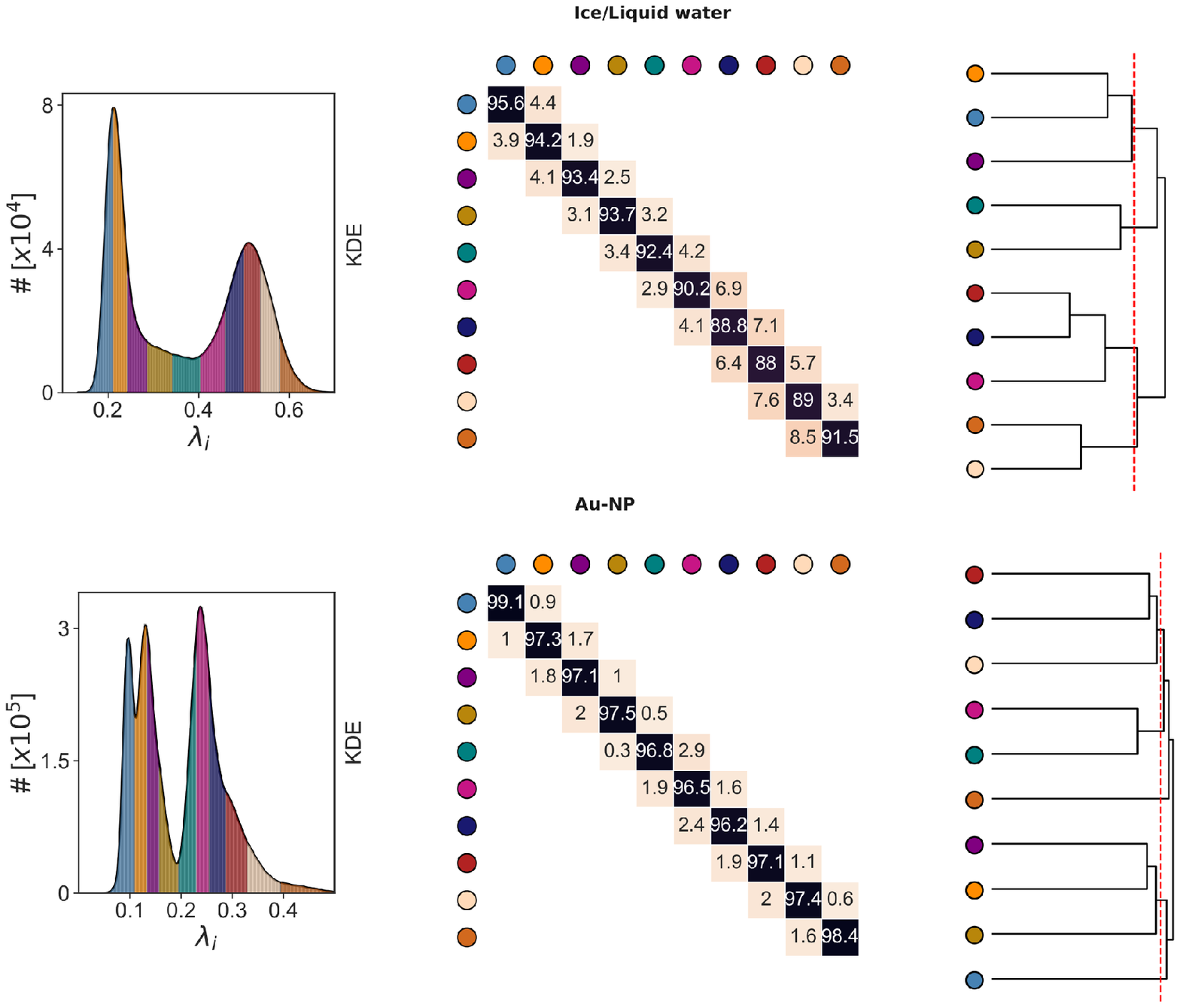}
\centering
\caption{Steps from the starting $K = 10$ clusters to the final $k$ clusters for Ice/Liquid water system (top) and Au-NP system (bottom). \textit{$\tau$}SOAP distribution with the 10 clusters identified by KMeans, the related exchange probability matrix and the hierarchical dendrogram with the cut-off are reported for each system.}
\label{SIfig03}
\end{figure*}

\begin{figure*}[ht!]
\includegraphics[width=18cm]{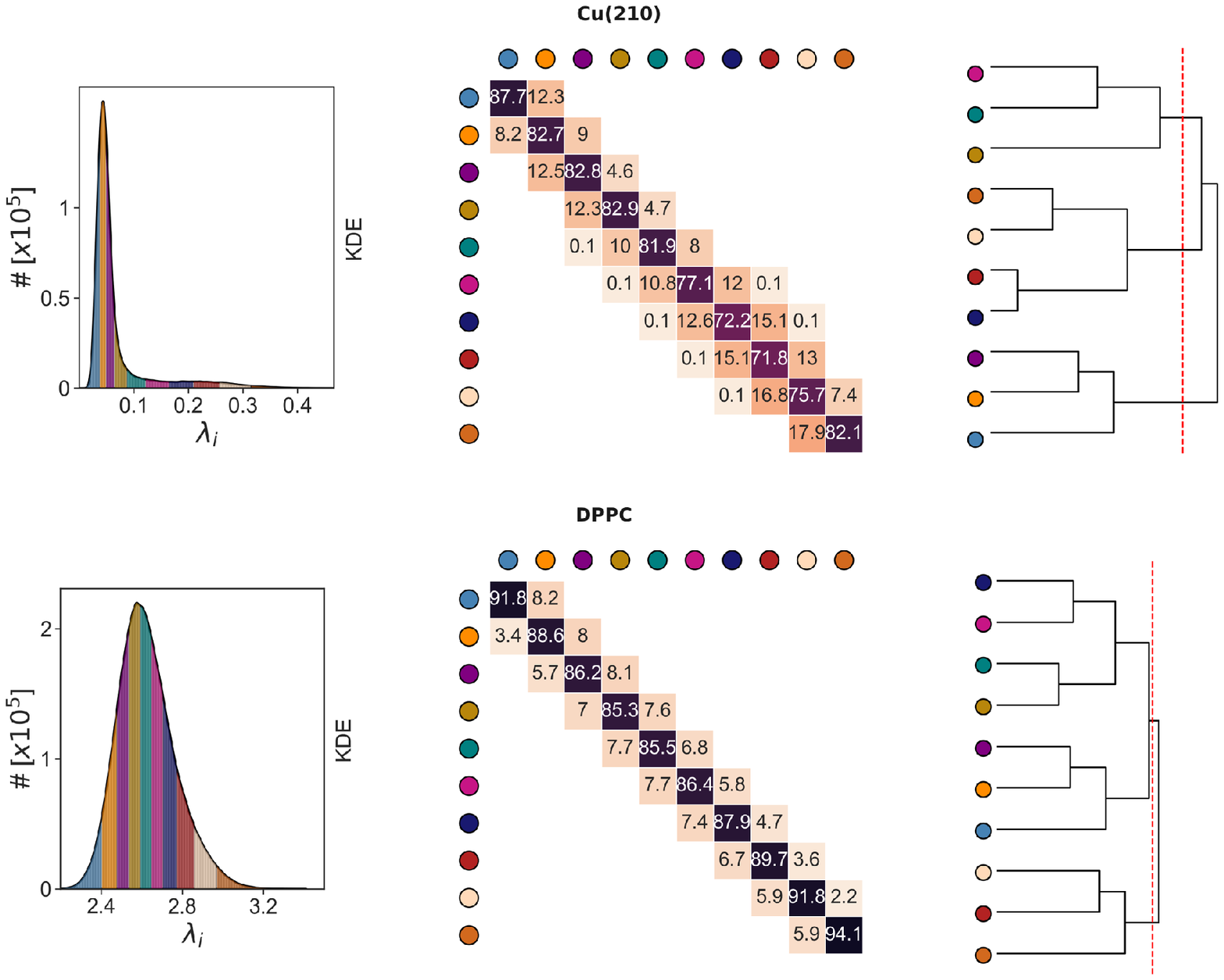}
\centering
\caption{Steps from the starting $K = 10$ clusters to the final $k$ clusters for Cu(210) (top) and DPPC lipid (bottom). \textit{$\tau$}SOAP distribution with the 10 clusters identified by KMeans, the related exchange probability matrix and the hierarchical dendrogram with the cut-off are reported for each system.}
\label{SIfig04}
\end{figure*}
\clearpage
\bibliography{bibliography}